\documentclass[aps,pra,twocolumn,groupedaddress,showpacs]{revtex4}

\usepackage{graphicx}
\usepackage{dcolumn}
\usepackage{bm}
\usepackage{verbatim}
\usepackage{amsmath,amssymb}
\usepackage{relsize,exscale}
\usepackage{color}
\begin{document}

\title{Hybrid quantum-classical models as constrained quantum systems }

\author{M. Radonji\'c}
\author{S. Prvanovi\'c}
\author{N. Buri\'c}
\email[]{buric@ipb.ac.rs}
\affiliation{Institute of Physics, University of Belgrade,
Pregrevica 118, 11080 Belgrade, Serbia}

\begin{abstract}
Constrained Hamiltonian description of the classical limit is utilized
in order to derive consistent dynamical equations for hybrid quantum-classical
systems. Starting with a compound quantum system in the Hamiltonian
formulation conditions for classical behavior are imposed on one of its
subsystems and the corresponding hybrid dynamical equations are derived.
The presented formalism suggests that the hybrid systems
have properties that are not exhausted by those of quantum and classical systems.
\end{abstract}

\maketitle

\subsection{ Introduction}
\vspace{-3mm}

Fundamental assumption of quantum mechanics is that the evolution of an isolated
quantum system is given by the linear Schr\"oedinger equation. On the other hand,
all macroscopic systems usually obey nonlinear evolution equations of classical
mechanics to an excellent approximation. The classical and the quantum theory
have developed different formalism to successfully describe interactions between
systems belonging to their respective domains. Correlations between quantum
objects are mathematically captured by the direct product structure of the Hilbert
spaces. On the other hand compound classical systems are described on the
Cartesian product of the component's phase spaces. Attempts to formulate a
consistent dynamical theory of interacting quantum-classical, commonly called hybrid,
systems are numerous as is illustrated by the following rather partial list of
references [1-8]. Current technologies are sufficiently developed to enable
experimental studies of the interaction between typically quantum and typically
classical objects \cite{Exper1,Exper2}, but such experiments require detailed
preliminary theoretical models.

In this work the framework of the theory of Hamiltonian dynamical systems is used
to treat the hybrid quantum-classical systems and to develop a description of
the interactions within such systems which is consistent with the main physically
justified requirements. In fact, it is well known [6,11-17] that quantum mechanics
can be formalized as a Hamiltonian dynamical system with the corresponding phase space
and with the quantum observables described by functions which are quadratic forms
of the canonical variables. More general functions on the quantum phase space do
not have any physical interpretation. This formalism is used in \cite{Else}
to develop a description of the hybrid classical-quantum systems by treating both,
quantum and classical, formally as Hamiltonian systems described in the Hamiltonian
language. The coupling between the systems is introduced somewhat {\it ad hoc }
as if both systems were classical, just because they are both described in the
framework of the Hamiltonian dynamical systems. This assumption about the treatment
of compound systems is not trivially obvious. For example, such treatment of
coupling between two quantum systems, both separately described in the Hamiltonian
framework, would be incorrect. In this paper we start with the total compound
quantum system in the geometric Hamiltonian framework. The next step is to
consider a classical limit of one of the component systems. To this purpose we
utilize our recently developed theory of general quantum constraints within the
Hamiltonian approach \cite{JaAnnPhys}, and the corresponding description of the
classical limit \cite{usPRA1,usPRA2}. The classical behavior of one of the
components is accomplished by constraining the Hamiltonian evolution so that the
quantum fluctuations of the would be classical degrees of freedom remain minimal
for all times. This in effect constrains the evolution onto a manifold which is
the Cartesian product of the quantum phase space of the quantum subsystem and the
manifold corresponding to the coherent states of the would be classical one.
The evolution equations for the interacting hybrid systems are obtained in the
macro limit applied on the coarse-grained subsystem and the constrained
Hamiltonian equations. The Hamiltonian form of the derived evolution equations
of the hybrid system turns out to be the same as the one postulated in Ref.\
\cite{Else} and therefore satisfies a list of standard requirements collected
and tested in \cite{Else}. We also provide a discussion of a puzzling fact,
pointed out also in \cite{Else}, regarding the physical interpretation of
functions of both the classical and the quantum degrees of freedom.

\vspace{-5mm}
\subsection{ Selective Coarse-graining and hybrid dynamics}
\vspace{-3mm}

The framework of constrained Hamiltonian description for the treatment of quantum
systems with nonlinear constraints and its application on the problem of classical
limit was developed and discussed in \cite{usPRA1,usPRA2} and shall not be repeated
here. Here we apply the general theory in order to derive consistent dynamical
equations for the hybrid systems. Consider a quantum system composed of two
quantum subsystems, for convenience fancifully called the first and the second.
The Hilbert space of the composite is ${\cal H}={\cal H}_1\otimes{\cal H}_2$ and
the quantum phase space of the composite is denoted ${\cal M}$. An arbitrary vector
from ${\cal H}$ is denoted $|\psi\rangle\!\rangle$ and the corresponding point
from ${\cal M}$ has complex canonical coordinates $(\psi(x),\psi^{*}(x))$ which
are expansion coefficients in a basis $\{|x\rangle\!\rangle\}$ of $|\psi\rangle\!\rangle$
and its dual vector. The Poisson bracket of a two functions on ${\cal M}$
is given by
\begin{equation}
\{f_1,f_2\}_{\cal M}=\frac{1}{i\hbar}\int dx\left(\frac{\delta f_1}{\delta \psi(x)}
\frac{\delta f_2}{\delta \psi^{*}(x)}-\frac{\delta f_2}{\delta \psi(x)}
\frac{\delta f_2}{\delta \psi^{*}(x)}\right).
\end{equation}
The notation for the basis $\{|x\rangle\!\rangle\}$ and the integral in (1) should be
understood symbolically and could denote a denumerable or finite basis and
summation respectively.
The Schr\"oedinger i.e.\ the Hamiltonian evolution is generated by the function
$H(\psi,\psi^{*})=\langle\!\langle\psi|\hat H_1+\hat H_2+\hat H_{12}|\psi\rangle\!\rangle$,
where the meaning of the notation is obvious.

For the sake of simpler presentation we shall consider the system such that the
first of the subsystems is in fact given by the $k-$fold product of the
Heisenberg algebras, that is by the basic operators $(\hat q_1,\dots,\hat q_k$,
$\hat p_1,\dots,\hat p_k)$ $\equiv(\hat q,\hat p)$.
We shall consider a hybrid classical-quantum system as a system such that the
the total quantum fluctuations of the first subsystem, that is the sum of
dispersions of the basic observables $ (\hat q,\hat p)$
\begin{equation}
F(X_{\psi})= \sum_{i=1}^k\big((\Delta\hat q_i)_{\psi}^2+(\Delta\hat p_i)_{\psi}^2\big)-min=0,
\end{equation}
is preserved minimal during the evolution. This condition represents a nonlinear
constraint on the admissible states of the total system. The evolution of the fully
quantum composite system must be modified in such a way that the constraint is
respected, and to this end we use the method developed in \cite{JaAnnPhys,usPRA1}.
The manifold $\bar\Gamma$ of the constraint (2) is a nonlinear symplectic
submanifold of ${\cal M}$ locally isomorphic with the Cartesian product
$\Gamma_1\times{\cal M}_2$, where $\Gamma_1$ is the manifold of the
 standard Heisenberg algebra minimal uncertainty coherent states (MUCS)
of the first subsystem, denoted by $|\alpha\rangle$ or $|q,p\rangle$, and
${\cal M}_2\sim {\cal H}_2$ is the quantum phase space of the second subsystem.
Therefore, at each point $|{\cal C}\rangle\!\rangle$ of $\bar\Gamma$ given by
\begin{equation}
|{\cal C}\rangle\!\rangle=|\alpha\rangle|\omega_2\rangle
\equiv|q,p\rangle|\omega_2\rangle,
\end{equation}
there are local symplectic coordinates $(q,p,\omega_2(x_2)$, $\omega_2^{*}(x_2))$
expressed in terms of $|{\cal C}\rangle\!\rangle$ as
$q=\langle\!\langle{\cal C}|\hat q|{\cal C}\rangle\!\rangle$,
$p=\langle\!\langle{\cal C}|\hat p|{\cal C}\rangle\!\rangle$
and $\omega_2(x_2)=\langle x_2|\langle q,p|{\cal C}\rangle\!\rangle$. The vectors
$\{|x_2\rangle\}$ symbolize a basis in ${\cal H}_2$ not necessarily the generalized
eigen-basis of some multiplication operator as the notation might suggest.
Notice that the requirement of minimal quantum fluctuations set only on
the first subsystem automatically implies that the first subsystem is always in
a coherent state and there is no entanglement between the two subsystems.
Also, there can be no entanglement between the degrees of freedom of the first
subsystem. No restriction on the type of states of the second subsystem is
set by the constraint (2), so the quantum subsystem can be in an entangled state.

The fundamental assumption concerning the dynamics of the putative hybrid system
is that the nonlinear constraint (2) is preserved during such evolution. This ensures
that the first subsystem is minimally quantum (as closest as possible to classical)
while the second subsystem is quantum in nature. Thus, our proposal for the dynamical equations
of these coupled subsystems are the Hamiltonian equations given by the original
Hamiltonian plus the additional terms that guarantee the preservation of the
constraint (2). The resulting equations will by construction preserve the
minimally quantum nature of the first subsystem.

The constrained manifold $\bar\Gamma$ is symplectic and in this case, as was
explained in detail in \cite{usPRA1,usPRA2}, the constrained system is Hamiltonian
with the Hamilton's function given by the original Hamilton's function
$\langle\!\langle\psi|\hat H|\psi\rangle\!\rangle$ evaluated on the constrained manifold.
Therefore, the dynamics is generated by the Poisson bracket on $\cal M$ and the Hamiltonian
\begin{eqnarray}
H_t&=&\langle\!\langle {\cal C}(\psi)|\hat H |{\cal C}(\psi)\rangle\!\rangle=
\langle\!\langle\psi|q,p\rangle\langle q,p|\hat H|q,p\rangle\langle q,p|\psi\rangle\!\rangle
\nonumber\\&\equiv&\langle\!\langle \psi|\hat H_{\alpha}(q,p)|\psi\rangle\!\rangle,
\end{eqnarray}
where $\hat H_{\alpha}(q,p)\equiv |q,p\rangle\langle q,p|\otimes \langle q,p|\hat H|q,p\rangle$.
In fact the constrained evolution of an arbitrary function-observable
$A(\psi)=\langle\!\langle\psi|\hat A|\psi\rangle\!\rangle$ on the constrained manifold
is obtained by reducing the following equation
\begin{eqnarray}\hspace*{-3mm}
\dot A(\psi)&=&\{A(\psi),H_t\}_{\cal M}\nonumber\\
&=&\frac{1}{i\hbar}\int dx \left(\frac{\delta A(\psi)}{\delta \psi(x)}\frac{\delta H_t}{\delta \psi^{*}(x)}-
\frac{\delta H_t}{\delta\psi(x)}\frac{\delta A(\psi)}{\delta\psi^{*}(x)}\right)
\end{eqnarray}
on the constrained manifold $\bar \Gamma$.

 For example, before reduction on $\bar \Gamma$ the dynamical equation for
$q=\langle\!\langle\psi|\hat q|\psi\rangle\!\rangle$ and
$ p=\langle\!\langle\psi|\hat p|\psi\rangle\!\rangle$ are given, by
\begin{subequations}
\begin{eqnarray}
\dot q&=&\frac{1}{i\hbar}\langle\!\langle\psi|[\hat q,\hat H_{\alpha}]|\psi\rangle\!\rangle
+\frac{\partial H_t}{\partial p},\\
\dot p&=&\frac{1}{i\hbar}\langle\!\langle\psi|[\hat p,\hat H_{\alpha}]|\psi\rangle\!\rangle
-\frac{\partial H_t}{\partial q}.
\end{eqnarray}
\end{subequations}
Short computation shows that the first terms in these equations are in fact equal
to zero on the constrained manifold $\bar\Gamma$. In fact, for an arbitrary operator
$\hat A_1$ acting only in ${\cal H}_1$ one has
$\langle\!\langle\psi|[\hat A_1,\hat H_{\alpha}]|\psi\rangle\!\rangle|_{\bar\Gamma}=0$.
Therefore, the dynamical equations for the first system's coordinates and momenta are
\begin{equation}
\dot q=\frac{\partial H_t}{\partial p}\quad
\dot p=-\frac{\partial H_t}{\partial q}.
 \end{equation}

Let us now compute the dynamical equations for the functions of the form
\begin{equation}
\omega_2(x_2)\equiv\langle x_2|\omega_2(\psi)\rangle=
\langle x_2|\langle q,p|\psi\rangle\!\rangle.
\end{equation}
Starting again with the equation
\begin{equation}
\dot \omega_2(x_2)=\frac{1}{i\hbar}\int dx \left(
\frac{\delta\omega_2}{\delta \psi(x)}\frac{\delta H_t}{\delta \psi^{*}(x)}-
\frac{\delta H_t}{\delta\psi(x)}\frac{\delta\omega_2}{\delta\psi^{*}(x)}\right)
\end{equation}
and after somewhat lengthy calculation one obtains before the reduction on
$\bar \Gamma$
\begin{eqnarray}\hspace*{-5mm}
i\hbar\,\dot \omega_2(x_2)&=&\langle x_2|\langle q,p|\hat H|q,p\rangle|\omega_2\rangle
\nonumber\\&+&\left(\frac{q}{2}\frac{\partial H_t}{\partial q}
+\frac{p}{2}\frac{\partial H_t}{\partial p}\right)\omega_2(x_2)\nonumber\\
&+&\frac{i}{\hbar}\langle x_2|\langle q,p|(\hat p-p/2)|\psi\rangle\!\rangle
\langle\!\langle \psi|[\hat q,\hat H_{\alpha}]|\psi\rangle\!\rangle\nonumber\\
&-&\frac{i}{\hbar}\langle x_2|\langle q,p|(\hat q-q/2)|\psi\rangle\!\rangle
\langle\!\langle \psi|[\hat p,\hat H_{\alpha}]|\psi\rangle\!\rangle.
\end{eqnarray}
Upon reduction on the constrained manifold $\bar\Gamma$ the last two terms are
annulled and the relevant dynamical equations can be written in the form
\begin{equation}
i\hbar\,\dot\omega_2(x_2)=\langle x_2|\langle \alpha|\hat H|\alpha\rangle|\omega_2\rangle
+\left(\frac{q}{2}\frac{\partial H_t}{\partial q}+
\frac{p}{2}\frac{\partial H_t}{\partial p}\right)\omega_2(x_2).
\end{equation}
The last term of this equation implies pure phase change and can be gauged away
resulting with
\begin{equation}
i\hbar\,\dot\omega_2(x_2;\psi)=\langle x_2|\langle \alpha(\psi)|
\hat H|\alpha(\psi)\rangle|\omega_2(\psi)\rangle.
\end{equation}
The equation (12) has the form of a Schr\"oedinger equation for the state vector
$\omega_2(x_2;\psi)=\langle x_2|\langle q,p|\psi\rangle\!\rangle\in {\cal H}_2$,
with the Hamiltonian operator $\langle \alpha(\psi)|\hat H|\alpha(\psi)\rangle$
acting on ${\cal H}_2$ and depending on
$q=\langle\!\langle\psi|\hat q|\psi\rangle\!\rangle$ and
$p=\langle\!\langle\psi|\hat p|\psi\rangle\!\rangle$.


The dynamical equations (7) and (12) can be written as Hamiltonian dynamical
equations in local coordinates on the constrained manifold $\bar\Gamma$ by
introducing the Poisson bracket on $\bar\Gamma$ for arbitrary functions on
$\bar\Gamma$ represented in the local coordinates $(q,p,\omega_2,\omega^{*}_2)$ as
\begin{eqnarray}
\hspace{-9mm}&&\{f_1,f_2\}_{\bar\Gamma}=\sum_{i=1}^k\left(\frac{\partial f_1}{\partial q_i}
\frac{\partial f_2}{\partial p_i}-\frac{\partial f_2}{\partial q_i}
\frac{\partial f_1}{\partial p_i}\right )\nonumber\\
\hspace{-9mm}&&+\frac{1}{i\hbar}\!\int\! dx_2\!\left( \frac{\delta f_1}{\delta \omega_2(x_2)}
\frac{\delta f_2}{\delta \omega_2^{*}(x_2)}-\frac{\delta f_2}{\delta \omega_2(x_2)}
\frac{\delta f_1}{\delta \omega_2^{*}(x_2)}\right)\!.
\end{eqnarray}
Thus, the Hamiltonian form of the hybrid dynamics on the constrained manifold
$\bar\Gamma$ as the phase space reeds
\begin{eqnarray}
&&\dot q=\{q,H_t\}_{\bar \Gamma},\quad \dot p=\{p,H_t\}_{\bar \Gamma},\\
&&\dot\omega_2=\{\omega_2,H_t\}_{\bar \Gamma},\quad \dot\omega^{*}_2
=\{\omega^{*}_2,H_t\}_{\bar \Gamma},
\end{eqnarray}
where the Hamilton's function $H_t(q,p,\omega_2(x_2),\omega^{*}_2(x_2))$ in local
coordinates on $\bar\Gamma$ is given by (4).


At this point we may briefly discuss the case of general i.e.\ mixed quantum states.
Such a state is given by a positive normalized function on ${\cal M}$ which has
quadratic dependence on the canonical coordinates $(\psi,\psi^{*})$.
Restriction of such a function on the nonlinear submanifold $\bar \Gamma$
results with a function $\rho(q,p,\omega_2(x_2),\omega_2^{*}(x_2))$ which depends
quadratic on $(\omega_2(x_2), \omega_2^{*}(x_2))$. On the other hand,
$\rho(q,p,\omega_2(x_2),\omega_2^{*}(x_2))$ for fixed $(\omega_2(x_2),\omega_2^{*}(x_2))$
can be an arbitrary positive function of $(q,p)$ with a unit integral over $\Gamma$,
since $(q,p)$ are not a subset of the canonical coordinates on ${\cal M}$, but
are physical observables $q=\langle\!\langle \psi |\hat q|\psi\rangle\!\rangle$,
$p=\langle\!\langle \psi |\hat p|\psi\rangle\!\rangle$. In fact they are a subset
of the canonical coordinates on the nonlinear submanifold $\bar\Gamma$. In terms
of the Poisson bracket on $\bar\Gamma$ the dynamics of $\rho(q,p,\omega_2(x_2),\omega_2^{*}(x_2))$
is given by the corresponding Liouville equation
\begin{equation}
\dot\rho(q,p,\omega_2,\omega_2^{*})=\{H_t(q,p,\omega_2,\omega_2^{*}),
\rho(q,p,\omega_2,\omega_2^{*})\}_{\bar \Gamma}.
\end{equation}

The constrained dynamics which preserves minimal value of the quantum fluctuations
of one of the subsystems is only the first step. The second step is the relevant
macro-limit so that the minimal quantum fluctuations still present in the
corresponding coherent states can be neglected when compared with actual values
of the dynamical variables. Therefore the macro-limit should be applied on the
equations (14) relevant for the first subsystem. This is illustrated in the
following example.

\vspace{-4mm}
\section{ An example: Two $1/2$-spins and a classical nonlinear oscillator}
\vspace{-3mm}

Consider a system of interacting equal qubits each coupled to the same nonlinear
oscillator. The quantum Hamiltonian of the total system is
\begin{equation}
\hat H= \varepsilon\hat\sigma_1^z+\varepsilon\hat\sigma_2^z+\mu\hat\sigma_1^x\hat\sigma_2^x
+\frac{\hat p^2 }{2m}+V(\hat q)+\hat q(\lambda_1\hat \sigma_1^z+ \lambda_2\hat \sigma_2^z),
\end{equation}
where $V(\hat q)$ is a polynomial expression in terms of $\hat q$ such that
$d^2 V(q)/dq^2|_{q=0}=m\Omega^2$. The constraining and the macro-limit will be applied on the
nonlinear oscillator subsystem.

The total Hamilton function of the constrained system is
$H_t=\langle\!\langle{\cal C}(\psi)|\hat H|{\cal C}(\psi)\rangle\!\rangle$, where
$|{\cal C}(\psi)\rangle\!\rangle=|q,p\rangle|\omega\rangle$. The complex coefficients
of an arbitrary $\omega\in\mathbb{C}^4$ in the computational basis are denoted by
$c_1$, $c_2$, $c_3$, $c_4$ and their real and imaginary components are the canonical
coordinates given by $(x_i,y_i)=\sqrt{2}({\rm Re}(c_i),{\rm Im}(c_i))$, $i=1,2,3,4$.
The expectation of the spin part of the Hamiltonian (17) in the vector
$|{\cal C}\rangle\!\rangle$ is
\begin{equation}
H_s=\varepsilon(y_1^2+x_1^2-y_4^2-x_4^2)+\mu(y_2y_3+y_1y_4+x_2x_3+x_1x_4).
\end{equation}
 he expectation in a vector $|{\cal C}\rangle\!\rangle$ of the interaction part is
\begin{eqnarray}
&&H_{int}=\lambda_1 q(y_1^2+y_2^2-y_3^2-y_4^2+x_1^2+x_2^2-x_3^2-x_4^2)/2\nonumber\\
&&+\lambda_2 q(y_1^2-y_2^2+y_3^2-y_4^2+x_1^2-x_2^2+x_3^2-x_4^2)/2,
\end{eqnarray}
 where $q=\langle q,p|\hat q|q,p\rangle$ is the coherent state expectation of the
oscillator's coordinate. Finally, the $|{\cal C}\rangle\!\rangle$ expectation of
the oscillator's Hamiltonian is
\begin{equation}
H_{osc}=\frac{p^2}{2m}+V(q)+\sum_{k=1}^{\infty}\frac{1}{2^k k!}\frac{\hbar^k V^{(2k)}(q)}{(2m\Omega)^k},
\end{equation}
where we used the explicit expression of $\langle q,p|V(\hat q)|q,p\rangle$
derived in \cite{usPRA1}. In the macro-limit the term containing
$\hbar\rightarrow 0$ is zero, leading to the Hamiltonian of the classical
nonlinear oscillator.

The total Hamiltonian generating the dynamics of the five degrees of freedom
$(q,p)$ and $(x_i,y_i)$, $i=1,2,3,4$ via the Hamiltonian dynamical equations (14) and (15)
is the sum of the three functions (18), (19) and (20).

The dynamics of the two qubits in the form of the Schr\"oedinger equation (12) is
given by the Hamilton operator on ${\cal H}_2=\mathbb{C}^4$ which depends also on
the oscillator coordinate $q=\langle q,p|\hat q|q,p\rangle$
\begin{equation}
\langle q,p |\hat H|q,p\rangle=\varepsilon \hat\sigma_z^1+
\varepsilon\hat\sigma_z^2+\mu\hat\sigma_x^1\hat\sigma_x^2+
\lambda_1 q\hat\sigma_z^1+\lambda_2 q\hat\sigma_z^1.
\end{equation}

\vspace{-5mm}
\section{ Discussion and summary}
\vspace{-3mm}

In summary, we have derived from the first principles the Hamiltonian dynamical
model corresponding to the hybrid quantum-classical systems that has been
postulated in \cite{Else}. In the derivation we have started from a quantum system
composed of two quantum subsystems and then we have assumed that one of the
subsystems has and preserves the classical properties during the interaction with
the quantum subsystem. This is implemented by the corresponding constrained
Hamiltonian dynamics. In this way the approach adopted in \cite{Else} is
justified from the first principles, which is our main result.


The main properties of the hybrid dynamical equations in their Hamiltonian form
(14) and (15) have been studied in detail in \cite{Else} and therefore need not
to be repeated here. However, we would like to comment on the following peculiar
property of the hybrid Hamiltonian dynamical system already analyzed in \cite{Else}.
Consider the Liouville equation (16) in the case that the hybrid Hamiltonian
$H_t(q,p,\omega_2,\omega_2^{*})$ and the density $\rho(q,p,\omega_2,\omega_2^{*})$
both depend on the same canonical pair of the classical subsystem. If the density
$\rho (q,p,\omega_2,\omega_2^{*})$ generates a mixed state on the quantum subsystem
then it must be a quadratic function of the canonical coordinates
$(\omega_2(x_2),\omega_2^{*}(x_2))$ corresponding to the quantum subsystem. The
Hamiltonian is also a quadratic function of the canonical coordinates of the
quantum subsystem. However, the set of such quadratic functions of the quantum
canonical coordinates which also depend on the classical coordinates is not
closed under the Poisson bracket (13) on $\bar\Gamma$. This is in sharp contrast
with the purely quantum case. Therefore, the Hamiltonian dynamical model that
corresponds to the hybrid system must include functions of the quantum canonical
variables which do not have the physical interpretation of quantum observables.
In fact, the hybrid Hamiltonian dynamical system does not preserve the metrical
properties of the hybrid phase space $\bar\Gamma$. This is akin to the purely
classical case where the corresponding dynamics preserves the symplectic structure,
i.e. the system is Hamiltonian, but does not preserve the metrical properties,
which are therefore not considered as a part of the classical system's structure.
Analogously, hybrid mixed states, i.e.\ probability densities on $\bar\Gamma$,
must be assumed to be of a more general form than in the purely quantum case.
Quantum mechanical average of an observable $\hat F$ in the state
$\hat\rho$: $\bar F={\rm Tr}[\hat\rho\hat F]$ is reproduced with
$\bar F=\int_{\cal M} F(X)\mu(X)dX$ using any of the probability densities
$\mu(X)$ with the same first moment that is fixed by the requirement that the
quantum expectation is equal to $\bar F$. The fact that the quantum mixed state
$\hat\rho$ determines only an equivalence class of densities $\mu(X)$, those with
the appropriate first moment, is equivalent to the non-uniqueness of the expansion
of the mixed state in terms of convex combinations of pure states and is crucially
quantum property of the Hamiltonian system on ${\cal M}$. We see that in the hybrid
systems even if the initial state $\rho (q,p,\omega_2,\omega_2^{*})$ generates a
quantum mixed state on the quantum subsystem, i.e.\ is quadratic in terms of the
canonical variables of the quantum subsystem, such a state will evolve into a
probability density of a more general form. This fact suggest that the truly hybrid
systems, if existent, must be considered as conceptually independent class and
not as such whose properties are exhausted by the properties of quantum and of
classical systems.

\acknowledgments

This work was supported in part by the Ministry of Science and Education of the
Republic of Serbia, contracts No.\ 171017, 171028, 171038, 171006 and 45016.
and by COST (Action MP1006).

\end{document}